\documentclass[a4paper]{article}
\usepackage{epsfig}
\usepackage{amsmath}
\newcommand\email[1]{}
\newcommand\institution[1]{}
\newcommand\address[1]{}
\begin{document}
\begin{center}
{\large {\bf On statistical methods of structure function extraction}}

\vspace{5mm}

S.N. Sevbitov$^{\dag}$\footnote{{serg\_sevbitov@tut.by}}, T.V. Shishkina$^{\dag}$\footnote{{shishkina@bsu.by}} and I.L. Solovtsov$^\ddag$\\
\vspace{5mm}
$^{\dag}${\it Belarusian State University, Nezavisimosti Av. 4, Minsk, 220030, Belarus}\\
$^{\ddag}${\it Deceased, formerly at Gomel State Technical University, October Av. 48,  Gomel, 246746, Belarus}

\end{center}


\vskip 5mm

\abstract{Several methods of statistical analysis  are proposed and analyzed in application for a specific task -- extraction of the structure functions from the cross sections of deep inelastic interactions of any type. We formulate the method based on the orthogonal weight functions and on an optimization procedure of errors minimization as well as methods underlying common $\chi^2$ minimization. We analyze effectiveness of these methods usage by comparison of the statistical parameters such as bias, extraction variance etc., for sample deep inelastic scattering data set.}


\section*{Introduction}

Precise extraction and analysis of the nucleon's structure functions from the deep inelastic scattering experiments plays great role in current understanding of the particles structure and development of quantum field theory. Deep inelastic scattering (DIS) processes or probing internal structure of nucleon by a pointlike particle at small distances and high energies provide such information about nucleon structure. As soon as experimental facilities to investigate polarized particle interactions have appeared, interest moved mainly to the study of the polarized DIS that allows us to inquire into the structure of nucleon's spin. Phenomenologically obtained cross sections and asymmetries contain such information as polarized structure functions $g_1$ and $g_2$ (or sometimes called $g_5$ or $g_6$) or partonic composition of nucleon's spin. To extract these data it's necessary to count on radiative corrections and background effects for obtaining pure cross sections (so-called unfolding of radiative smearing, see e.g. \cite{hermes,zeus}) for further extraction of the structure functions. This paper is devoted to some statistical aspects in application to extraction of the structure functions from the Born interaction parameters.

The cross sections for neutral and charged current deep inelastic scattering both on unpolarized and polarized nucleon targets can be written in the following ``separable'' form (see e.g. \cite{pdg}):
\[d^2\sigma/dxdy\equiv\sigma_{xy}=\sigma_0 \sum_{k}{Y_k (y,x,Q^2)W_k(x,Q^2),} \]
where $Y_k(y,x,Q^2)$ are known functions, which in the Bjorken limit depend on $y$ only and they are $y$-polynomial. $W_k(x,Q^2)$ are hadronic structure functions. For massless leptons the cross section is parameterized by three unpolarized structure functions
\[W_k^{\rm unpolarized}(x,Q^2)\Rightarrow F_1(x,Q^2),\;F_2(x,Q^2),\;F_3(x,Q^2)\]
and five polarized functions
\[W_k^{\rm polarized}(x,Q^2)\Rightarrow g_1(x,Q^2),\;g_2(x,Q^2),\;g_3(x,Q^2),\;g_3(x,Q^2),\;g_3(x,Q^2),\;g_3(x,Q^2).\]

One can derive desired (un)polarized structure functions operating directly with asymmetries, nevertheless, as asymmetries and corresponding cross sections are being related with each other,  we propose here methods operating only with the cross sections. Procedures and methods described below can be easily generalized for more specific cross section expressions, e.g. extended phenomenological models with larger number of the structure functions or more complicated form of the $Y_k$ functions. These methods can be considered as an analogue of many-dimensional interpolation within a given hypothesis for mean values at some points of some random variable. Constraints on applicability are possible though.

The structure of the given article is the following: after initial experimental (statistical) data being introduced we sketch orthogonal weight function method and its optimization technique for the simplest case of the cross section that includes two structure functions $f(x)$ and $g(x)$ for methodological purpose (in the case of polarized particles DIS one can get such expressions by subtracting corresponding cross sections with opposite spin directions). Then we offer $\chi^2$ minimization methods adapted to a given task. Last section compares mentioned methods by means of numerical estimation.

\section*{Initial experimental data}
Consider normalized on a $\sigma_0$ the cross section of the simplest form
\begin{equation}\label{cs}
    \sigma_{xy}=Y_{+}(y)f(x) +Y_{-}(y)g(x),
\end{equation}
where $Y_{\pm}$ reads
\begin{equation}\label{Y}Y_{\pm}=1\pm(1-y)^2.\end{equation}
Let's suppose that we know values of cross sections $\sigma_{xy}$ (in other words -- a counted number of events in the bin of a histogram) and its errors $\Delta\sigma_{xy}$ at some $x,y$-lattice (grid) of experimental kinematical points. Defined grid can be either regular or irregular, so we will distinguish the following cases:
\begin{enumerate}
\item Regular (rectangular) $x$,$y$-grid
\[M=\mathbf{X}\otimes\mathbf{Y}, \quad \mathbf{X}=x_1,\ldots,x_m; \quad \mathbf{Y}=y_1,\ldots,y_n. \]
\item Regular (rectangular) $x$,$Q^2$-grid
\[\tilde{M}=\mathbf{X}\otimes\mathbf{Q^2}, \quad \quad \mathbf{X}=x_1,\ldots,x_m; \quad \mathbf{Q^2}={Q_1^2},\ldots,{Q_n^2}. \]
\item Non-regular (irregular) $x$,$y$- and $x$,$Q^2$-grid -- arbitrary set of points (bins).
\end{enumerate}

Although, it's not crucial here for the lattice to be regular (besides experimenters mostly gather events on irregular set of bins), but at first we implement methods for regular grids and then propose schemes to generalize and apply them for an irregular set of points (bins) using simple interpolation considerations.

\section*{Orthogonal weight function method}

Let's begin with the rectangular $x,y$-grid $M$ and take the parameters $a_{\pm}$ of some preliminary chosen weight function $\omega(y;a)$ in such a way that the following orthogonality condition fulfils:
\[\sum_y{\omega(y;a_{\pm})Y_{\pm}}=0,\]
whereas the following requirement holds:
\[\sum_y{\omega(y;a_{\mp})Y_{\pm}}\neq 0.\]
In another way, preceding expressions mean that $\omega$ acts as a projection operator.
Then one can extract structure function by projection in the following form:
\begin{equation}\label{owf}
\begin{split}
    f(x)\pm\Delta f(x)=\frac{1}{\sum_{y}{\omega(y;a_{-}) Y_{+}(y)}}{\sum_{y}{\omega(y;a_{-})[\sigma_{xy}\pm\Delta\sigma_{xy}]}},\\
    g(x)\pm\Delta g(x)=\frac{1}{\sum_{y}{\omega(y;a_{+}) Y_{-}(y)}}{\sum_{y}{\omega(y;a_{+})[\sigma_{xy}\pm\Delta\sigma_{xy}]}}.
\end{split}
\end{equation}

For example, one may choose the simplest weight functions as a $\omega(y;a)=1+ay$, then $a_\pm$ parameters take the following values:
\begin{equation}\label{a}a_\pm=-\frac{\sum_{y}{Y_{\pm}(y)} }{\sum_{y}{y\,Y_{\pm}(y)}}.\end{equation}
Note that in this case $a_\pm$ depend only on chosen data grid.

As presented above procedure of structure function extraction  implies rough estimation of the standard deviation (error) values in eq. (\ref{owf}).
To find correctly uncertainty in the fitted structure functions one should adhere to the standard procedure of the variance (noted here as $D$) calculation of random variables function, which assumes here for a given $x$-bin and a set of $n$ $y$-bins (the same holds for $g$-function)
\begin{equation}\label{Df}
\begin{split}
D[f(\sigma(x,y_1),\ldots,\sigma(x,y_n))]=\frac{\displaystyle 1}{\displaystyle\bigg[\sum_{i=1}^{n}{\omega(y_i;a_{-}) Y_{+}(y)}\bigg]^2}\times\bigg\{\sum_{i=1}^{n}{\omega(y_i;a_{-})^2 D[\sigma(x,y_i)]}+\\+\sum_{i\neq j}{\omega(y_i;a_{-})\omega(y_j;a_{-})\rho[\sigma(x,y_i),\sigma(x,y_j)]\sqrt{D[\sigma(x,y_i)]D[\sigma(x,y_j)]}}\bigg\}.
\end{split}
\end{equation}
In the case of the uncorrelated data with diagonal correlation matrix $\rho[\sigma(x,y_i),\sigma(x,y_j)]=\delta_{ij}$ we specify correct estimation for deviation of the structure function values
\begin{equation}\label{df}
\Delta f(\sigma(x,y_1),\ldots,\sigma(x,y_n))=\frac{1}{\displaystyle\left|\sum_{i=1}^{n}{\omega(y_i;a_{-}) Y_{+}(y_i)}\right|}\sqrt{\sum_{i=1}^{n}{\omega(y_i;a_{-})^2 D(\sigma(x,y_i))}},
\end{equation}
that can be larger or smaller than
\begin{equation}
\frac{1}{\displaystyle\sum_{i=1}^{n}{\omega(y_i;a_{-}) Y_{+}(y_i)}}{\sum_{i=1}^{n}{\omega(y_i;a_{-}) \Delta\sigma(x,y_i)}},\end{equation}
dependently of the $\omega(y;a)$ sign. In these formulas we assume that the lattice has no uncertainty, i.e. $D[a_{\pm}]=0$ as given initially.

\section*{Optimization procedure}

To minimize errors of the structure functions that are extracted from experimental data $\sigma_{xy}\pm\Delta\sigma_{xy}$ one can apply an optimization procedure of the following type. Let's introduce auxiliary functions
\[A(a)=\sum_{y}{\omega(y;a)Y_{+}(y)},\quad B(a)=\sum_{y}{\omega(y;a)Y_{-}(y)},\]
\[S(x;a)=\sum_{y}{\omega(y;a)\sigma_{xy}},\quad \Delta S(x;a)=\sum_{y}{\omega(y;a)\Delta\sigma_{xy}}.\]
Two systems of equations for $f(x)$ and $g(x)$ and errors $\Delta f(x)$ and $\Delta g(x)$ are
\[\left\{\begin{array}{lll}A(a)f(x)+B(a)g(x)=S(x;a),\hfill\\A(b)f(x)+B(b)g(x)=S(x;b);\hfill\end{array}\right.\]
\[\left\{\begin{array}{lll}A(a)\Delta f(x)+B(a)\Delta g(x)=\Delta S(x;a),\hfill\\A(b)\Delta f(x)+B(b)\Delta g(x)=\Delta S(x;b).\hfill\end{array}\right.\]
To find their solutions define the following determinants:
\[\Delta(a,b)=\left|\begin{array}{cc}\!A(a) \;\; B(a)\!\\\!A(b) \;\; B(b)\!\end{array}\right|,\quad \Delta_1(x;a,b)=\left|\begin{array}{cc}\!S(x;a) \;\; B(a)\!\\\!S(x;b) \;\; B(b)\!\end{array}\right|,\quad \Delta_2(x;a,b)=\left|\begin{array}{cc}\!A(a) \;\; S(x;a)\!\\\!A(b) \;\; S(x;b)\!\end{array}\right|\]
and
\[\delta\Delta_1(x;a,b)=\left|\begin{array}{cc}\!\Delta S(x;a) \;\; B(a)\!\\\!\Delta S(x;b) \;\; B(b)\!\end{array}\right|,\quad \delta\Delta_2(x;a,b)=\left|\begin{array}{cc}\!A(a) \;\; \Delta S(x;a)\!\\\!A(b) \;\; \Delta S(x;b)\!\end{array}\right|.\]
Then we get the solution
\[f(x)=\frac{\Delta_1(x;a,b)}{\Delta(a,b)}, \quad \Delta f(x)= \frac{\delta\Delta_1(x;a,b)}{\Delta(a,b)},\quad g(x)=\frac{\Delta_2(x;a,b)}{\Delta(a,b)}, \quad \Delta g(x)= \frac{\delta\Delta_2(x;a,b)}{\Delta(a,b)}.\]

The optimal values of the parameters $a$ and $b$ can be found from the condition of the errors minimization
\begin{equation}\label{min}\min_{\{a,b\}}[w_f|\Delta f(x)|+w_g|\Delta g(x)|],\end{equation}
where $w_f$ and $w_g$ -- optional weight factors.

So, optimization procedure implies determination of the optimal parameters $a_k$ and $b_k$ for each experimental point $x_k$ in order to minimize errors in this point. As a result we have the estimation for the mean values of the structure functions extracted at given experimental points and  rough estimations for corresponding errors (deviations):
\begin{equation}\label{opt}
f(x)=\frac{\Delta_1(x;a,b)}{\Delta(a,b)}, \quad \Delta f(x)= \frac{\delta\Delta_1(x;a,b)}{\Delta(a,b)},\quad g(x)=\frac{\Delta_2(x;a,b)}{\Delta(a,b)}, \quad \Delta g(x)= \frac{\delta\Delta_2(x;a,b)}{\Delta(a,b)},
\end{equation}
for each $k=1,2,\ldots, n$.

It should be noted that by such solution we get only approximate values for errors (likewise mentioned above argument about correct deviation values), nevertheless one can easily obtain correct values by finding minimum solution (\ref{min}) analytically and repeating formulas (\ref{Df}), (\ref{df}).


A difficulty arises from the fact that the method implies experimental data on a (rectangular) $x,y$-lattice which is rarely used, furthermore as a rule experimental bins chosen for analysis and fitting are not uniformly distributed (e.g. see kinematics in experimental reports    \cite{hermes,zeus}). First of all we propose to apply the same scheme to regular (rectangular) $x,Q^2$-lattice. The difference between $x,y$- and $x,Q^2$-data consist in principle only in redefinition of the structure functions. Let's modify $Y$ expressions (\ref{Y}):
\[
Y_{\pm}(Q^2)=1\pm\left(1-\frac{Q^2}{s\,x}\right)^2,\quad \tilde{Y}_{+}=Q^4,\quad \tilde{Y}_{-}=2Q^2,
\]
\[
{\sigma_{x,Q^2}}\sim 2\frac{f(x)}{x}+\frac{\tilde{f}(x)}{x}\frac{1}{s^2 x^2}{\tilde{Y}_{+}} - \frac{\tilde{f}(x)}{x}\frac{1}{s x}{\tilde{Y}_{-}}
 \]
and use the similar test weight function $\omega(Q^2;a)=1+a Q^2$. The same orthogonality relations take the form of
\[\sum_{Q^2}{\omega(Q^2;a_{\pm})\tilde{Y}_{\pm}}=0,\quad \sum_{Q^2}{\omega(Q^2;a_{\pm})}\neq 0,\]
\[a_{\pm}=-\frac{\sum_{Q^2}{\tilde{Y}_{\pm}}}{\sum_{Q^2}{Q^2\tilde{Y}_{\pm}}}.\]
As a result one can obtain
\[\sum_{Q^2}{\sigma_{x,Q^2}\omega(Q^2;a_{\pm})}=2\frac{f(x)}{x}\sum_{Q^2}{\omega(Q^2;a_{\pm})}+\frac{\tilde{f}(x)}{x}\frac{1}{s^2 x^2}\sum_{Q^2}{\omega(Q^2;a_{\pm})\tilde{Y}_{+}} - \frac{\tilde{f}(x)}{x}\frac{1}{s x}\sum_{Q^2}{\omega(Q^2;a_{\pm})\tilde{Y}_{-}},\]
\[\sum_{Q^2}{\sigma_{x,Q^2}\omega(Q^2;a_{+})}=2\frac{f(x)}{x}\sum_{Q^2}{\omega(Q^2;a_{+})} - \frac{\tilde{f}(x)}{s x^2}\sum_{Q^2}{\omega(Q^2;a_{+})\tilde{Y}_{-}},\]
\[\sum_{Q^2}{\sigma_{x,Q^2}\omega(Q^2;a_{-})}=2\frac{f(x)}{x}\sum_{Q^2}{\omega(Q^2;a_{-})}+\frac{\tilde{f}(x)}{s^2x^3}\sum_{Q^2}{\omega(Q^2;a_{-})\tilde{Y}_{+}},\]
where $\tilde{f}(x)=f(x)- g(x)$. Hence
\[f(x)=\frac{x}{2}\frac{\left[\sum_{Q^2}{\sigma_{x,Q^2}\omega(Q^2;a_{+})}\sum_{Q^2}{\omega(Q^2;a_{-})\tilde{Y}_{+}}+x s\sum_{Q^2}{\sigma_{x,Q^2}\omega(Q^2;a_{-})}\sum_{Q^2}{\omega(Q^2;a_{+})\tilde{Y}_{-}}\right]}{\left[\sum_{Q^2}{\omega(Q^2;a_{+})}\sum_{Q^2}{\omega(Q^2;a_{-})\tilde{Y}_{+}}+x s\sum_{Q^2}{\omega(Q^2;a_{-})}\sum_{Q^2}{\omega(Q^2;a_{+})\tilde{Y}_{-}}\right]},\]
\[g(x)=f(x)-x\frac{s^2 x^2\left[\sum_{Q^2}{\sigma_{x,Q^2}\omega(Q^2;a_{-})}\sum_{Q^2}{\omega(Q^2;a_{+})}- \sum_{Q^2}{\sigma_{x,Q^2}\omega(Q^2;a_{+})}\sum_{Q^2}{\omega(Q^2;a_{-})}\right]}{\left[\sum_{Q^2}{\omega(Q^2;a_{+})}\sum_{Q^2}{\omega(Q^2;a_{-})\tilde{Y}_{+}}+x s\sum_{Q^2}{\omega(Q^2;a_{-})}\sum_{Q^2}{\omega(Q^2;a_{+})\tilde{Y}_{-}}\right]}.\]
To obtain values of the absolute error one should substitute $\Delta \sigma_{x,Q^2}$ instead of $\sigma_{x,Q^2}$.

One of the advantages of $x,Q^2$-grid consists in possibility of appropriate usage of the interpolation through $Q^2$ range, contrary to the $y$ variable range, which combines different $Q^2=s x y$ values in an unhandy way. Special optimization procedure similar to one described above by formulas (\ref{opt}) is also applicable over here.
Used scheme will be reliable if one obtains data for rectangular $x,Q^2$-region with narrow $Q^2$ range to neglect existing $Q^2$-dependency of structure functions (but still with at least two different $Q^2$ values).
It should be noted that for non-uniform grids summing range $\sum_{Q^2}$ may depend on selected $x$-bin and consequently $a_{\pm}\rightarrow a_{\pm}(x)$ becomes a function of the given $x$ value. Nevertheless, above mentioned scheme works in the same way.

If one has $x,Q^2$-lattice of the data, which cannot be grouped in the certain $x$-bins, but experimental points are distributed in the vicinity of certain $x$-values, one can eliminate these difficulties by interpolation methods, e.g. as briefly mentioned below.

\subsection*{Interpolation}

Here we describe simplest ways to manage with data on irregular grids. One can use the following Taylor formula for the cross section value near given $x_i$ value
\[\sigma^{int}_i(x,Q^2)=\sigma^{\mathrm{ex}}(x_i,Q^2)+\frac{\sigma^{\mathrm{ex}}(x_i,Q^2)}{\sigma^{\mathrm{fit}}(x_i,Q^2)}(x-x_i)\partial_x{\sigma^{fit}(x_i,Q^2)}+\ldots\]
\[\sigma^{int}(x,Q^2)=[\text{weighted average}]=\frac{\sum_{i}{{\sigma^{int}_{i}(x,Q^2)}w(x-x_i)}}{\sum_{i}{w(x-x_i)}}.\]
Here the additional fraction included with the purpose of normalization - but it may be omitted. These formulas match the case when one has data with a small dispersion in $x$-values. One can employ some external parameterization, denoted here as $\sigma^{\mathrm fit}$, to fix missing experimental $x$ points by simple interpolation. But total uncertainty of the extracted $f(x)$ and $g(x)$ values increases by theoretical uncertainty of these parameterizations.

In the case of wide $Q^2$ range $Q^2$-dependence cannot be omitted and one can treat $f(x,Q^2)$ approximately using similar Taylor series of $f$ for $Q^2$ near given $Q_0^2$ value
\[f(x)\rightarrow f^{\mathrm{int}}(x,Q^2)=f(x,Q_0^2)+\frac{f(x,Q_0^2)}{f^{\mathrm{fit}}(x_i,Q_0^2)}(Q^2-Q_0^2)\partial_{Q^2} f^{\mathrm{fit}}+\ldots,\] or introducing some fixing factor $\delta(x,Q^2)$ of a pregiven form, e.g. $f(x)\delta(x,Q^2)$, but these ideas require $f(x)$ and $g(x)$ as well as $Y_{\pm}$ and $\tilde{Y}_{\pm}$ to be redesignated.

\subsection*{$\chi^2$-minimization procedure for the best $a_{\pm}$ estimations}

Below common $\chi^2$ methods are applied to problem of structure functions extraction. First we preserve usage of the $a_{\pm}$ parameters (thus implicit method, requiring introducing of the orthogonal function $\omega$) and modify common procedure. Let's construct the $\chi^2$-function  of random cross-sections and parameters $a_{-},a_{+}$ as follows
\[\chi^2(\sigma(x,y_1),\ldots,\sigma(x,y_n);a_{-},a_{+})=\sum_{i,j=1}^{n}{[\sigma(x,y_i) - m_i(a_{-},a_{+})]{D^{-1}}_{ij}[\sigma(x,y_j) - m_j(a_{-},a_{+})]},\]
where $D_{ij}=\rho[\sigma(x,y_i),\sigma(x,y_j)]\sqrt{D[\sigma(x,y_i)]D[\sigma(x,y_j)]}$ is the covariance matrix and ${m}(a_{-},a_{+})$ are the expectation values:
\[ {m}_i(a_{-},a_{+})=Y_{+}(y_i)\frac{\displaystyle\sum_{j=1}^{n}{\omega(y_j;a_{-}) \sigma(x,y_j)}}{\displaystyle\sum_{j=1}^{n}{\omega(y_j;a_{-}) Y_{+}(y_j)}} + Y_{-}(y_i)\frac{\displaystyle\sum_{j=1}^{n}{\omega(y_j;a_{+}) \sigma(x,y_j)}}{\displaystyle\sum_{j=1}^{n}{\omega(y_j;a_{+}) Y_{-}(y_j)}}.\]

As $a_{\pm}$ are supposed to be estimated parameters only, we may neglect dependence of these mean values on the random cross-section values $\sigma(x,y_j)$ regarding them as a set of initial exact numbers.

Next step is to minimize the  $\chi^2$ function. In the case of the uncorrelated data with $\rho[\sigma(x,y_i),\sigma(x,y_j)]=\delta_{ij}$ we get the following estimator equations:
\[
-\sum_{i=1}^{n}{\frac{\partial m_i}{\partial \hat{a}_{-}}D_{ii}^{-1}[\sigma(x,y_i) - m_i(\hat{a}_{-},\hat{a}_{+})]}=0,
\]
\[
-\sum_{i=1}^{n}{\frac{\partial m_i}{\partial \hat{a}_{+}}D_{ii}^{-1}[\sigma(x,y_i) - m_i(\hat{a}_{-},\hat{a}_{+})]}=0.
\]
Consistent solution of the system gives $\chi^2$-estimators $\hat{a}_{\pm}$. One can check consistency calculating the derivatives, which should not vanish
\[
-\sum_{i=1}^{n}{\frac{\partial^2 m_i}{\partial \hat{a}_{\pm}^2}D_{ii}^{-1}[\sigma(x,y_i) - m_i(\hat{a}_{-},\hat{a}_{+})]}+\sum_{i=1}^{n}{\Big(\frac{\partial m_i}{\partial \hat{a}_{\pm}}\Big)^2 D_{ii}^{-1}}\neq 0,\]\[
-\sum_{i=1}^{n}{\frac{\partial^2 m_i}{\partial \hat{a}_{+}\partial \hat{a}_{-}}D_{ii}^{-1}[\sigma(x,y_i) - m_i(\hat{a}_{-},\hat{a}_{+})]}+\sum_{i=1}^{n}{\frac{\partial m_i}{\partial \hat{a}_{-}}D_{ii}^{-1}\frac{\partial m_i}{\partial \hat{a}_{+}}}\neq 0.
\]
After that one should substitute obtained $\hat{a}_{\pm}$ into the following expression to get $f(x)$ and $g(x)$:
\[
f(x)=\frac{\displaystyle\sum_{j=1}^{n}{\omega(y_j;\hat{a}_{-}) \sigma(x,y_j)}}{\displaystyle\sum_{j=1}^{n}{\omega(y_j;\hat{a}_{-}) Y_{+}(y_j)}},\quad g(x)=\frac{\displaystyle\sum_{j=1}^{n}{\omega(y_j;\hat{a}_{+}) \sigma(x,y_j)}}{\displaystyle\sum_{j=1}^{n}{\omega(y_j;\hat{a}_{+}) Y_{-}(y_j)}}.
\]
Apart from these values, it's necessary to get estimations for the statistical parameters such as bias, deviation etc. This analysis is discussed below.

\subsection*{$\chi^2$-minimization procedure for the best $f(x)$ and $g(x)$ estimations}

Here we omit $a_{\pm}$ parameter and estimate functions $f(x)$ and $g(x)$ explicitly (as before we refer to $x$ as to fixed bin), thus changing  $\chi^2$ arguments:
\[\chi^2(\sigma(x,y_1),\ldots,\sigma(x,y_n);f(x),g(x))=\sum_{i,j=1}^{n}{[\sigma(x,y_i) - m_i(f(x),g(x))]{D^{-1}}_{ij}[\sigma(x,y_j) - m_j(f(x),g(x))]},\]
where $D_{ij}=\rho[\sigma(x,y_i),\sigma(x,y_j)]\sqrt{D[\sigma(x,y_i)]D[\sigma(x,y_j)]}$ is the covariance matrix and ${m}(f(x),g(x))$ are the expectation values:
\[ {m}_i(f(x),g(x))=Y_{+}(y_i)f(x) + Y_{-}(y_i)g(x).\]

Next step is to minimize the $\chi^2$ function. In the case of the uncorrelated data with diagonal correlation matrix $\rho[\sigma(x,y_i),\sigma(x,y_j)]=\delta_{ij}$ we get the following estimator equations:
\[
-\sum_{i=1}^{n}{Y_{+}(y_i) D_{ii}^{-1}[\sigma(x,y_i) - Y_{+}(y_i)f(x) - Y_{-}(y_i)g(x)]}=0,
\]
\[
-\sum_{i=1}^{n}{ Y_{-}(y_i)D_{ii}^{-1}[\sigma(x,y_i)- Y_{+}(y_i)f(x) - Y_{-}(y_i)g(x)]}=0.
\]
Solving this equations we get the following consistent $\chi^2$-estimates $f(x)$ and $g(x)$
\[\Delta=\left|\begin{array}{cc} \sum_{i=1}^{n}{Y_{+}^2(y_i)D_{ii}^{-1}} & \sum_{i=1}^{n}{Y_{+}(y_i)Y_{-}(y_i)D_{ii}^{-1}}\\\sum_{i=1}^{n}{Y_{+}(y_i)Y_{-}(y_i)D_{ii}^{-1}} & \sum_{i=1}^{n}{Y_{-}^2(y_i)D_{ii}^{-1}}\end{array}\right| \neq 0,
\]
\begin{equation}
\begin{split}\label{chif}
f(x)=\frac{1}{\Delta}\left|\begin{array}{cc} \sum_{i=1}^{n}{Y_{+}(y_i)D_{ii}^{-1}\sigma(x,y_i)} & \sum_{i=1}^{n}{Y_{+}(y_i)Y_{-}(y_i)D_{ii}^{-1}}\\\sum_{i=1}^{n}{Y_{-}(y_i)D_{ii}^{-1}\sigma(x,y_i)} & \sum_{i=1}^{n}{Y_{-}^2(y_i)D_{ii}^{-1}}\end{array}\right|,\\
g(x)=\frac{1}{\Delta}\left|\begin{array}{cc} \sum_{i=1}^{n}{Y_{+}^2(y_i)D_{ii}^{-1}} & \sum_{i=1}^{n}{Y_{+}(y_i)D_{ii}^{-1}\sigma(x,y_i)} \\\sum_{i=1}^{n}{Y_{+}(y_i)Y_{-}(y_i)D_{ii}^{-1}} & \sum_{i=1}^{n}{Y_{-}(y_i)D_{ii}^{-1}\sigma(x,y_i)}\end{array}\right|.
\end{split}
\end{equation}
One can check consistency calculating the derivatives, which should not vanish, indeed
\[
\sum_{i=1}^{n}{Y_{\pm}^2(y_i) D_{ii}^{-1}}\neq 0,\quad
\sum_{i=1}^{n}{Y_{+}(y_i) Y_{-}(y_i) D_{ii}^{-1}}\neq 0.
\]
Last method resembles previous optimization procedure (even in the forms of the matrices), but crucial point here is the absence of $a$ parameter (it looks like $Y$ is a weight function itself without any auxiliary parameter). For the case of correlated data one can refer to the same formulas using simple substitutions, e.g. in the matrix form:
\[
\sum_{i=1}^{n}{Y_{\pm}^2(y_i) D_{ii}^{-1}}\Rightarrow \sum_{i,j=1}^{n}{Y_{\pm}(y_i) D_{ij}^{-1}Y_{\pm}(y_j)}={\bf Y}_{\pm}^{T}D^{-1}{\bf Y}_{\pm}.
\]
Considered case corresponds to the linear $\chi^2$ approach, so using standard statistics one can get the following, e.g. \cite{stat}:
\begin{equation}\label{Dfg}
\begin{split}
D(f)=\frac{1}{\Delta^2}F D F^{T}, \quad F = \left({\bf Y}_{-}^{T}D^{-1}{\bf Y}_{-}\right){\bf Y}_{+}^{T}D^{-1}-\left({\bf Y}_{+}^{T}D^{-1}{\bf Y}_{-}\right){\bf Y}_{-}^{T}D^{-1},\\
D(g)=\frac{1}{\Delta^2}G D G^{T}, \quad G = \left({\bf Y}_{+}^{T}D^{-1}{\bf Y}_{+}\right){\bf Y}_{-}^{T}D^{-1}-\left({\bf Y}_{-}^{T}D^{-1}{\bf Y}_{+}\right){\bf Y}_{+}^{T}D^{-1}.
\end{split}
\end{equation}
The similar analytical estimation of the variance holds true for previous section, but main difference is that one can treat values $\sigma(x,y_j)$ either as fixed initial numbers or random ones with statistical uncertainty, what leads to additional complication of the (\ref{Dfg}) calculation (i.e. $m(a_{-},a_{+})$ holds itself error) and slightly increases the total deviations $\Delta f(x)$ and $\Delta g(x)$. Main advantage of the last $\chi^2$ approach (least squares method in the linear case) is that it provides consistent unbiased estimator for $f$ and $g$ with the smallest in its type estimator variance, according to the Gauss-Markov theorem \cite{stat}.
\section*{Discussion and numerical results}

For the purpose of numerical analysis let's take the model parameterization for the structure functions, e.g. in the following simplest form:
\begin{equation}
f(x)=C_f x^{\alpha_f}(1-x)^{\beta_f}, \quad g(x)=C_g x^{\alpha_g}(1-x)^{\beta_g}.
\end{equation}
Then one can construct cross section $\sigma_{xy}$ and its error $\Delta \sigma_{xy}$ at some points $x_1,\ldots,x_n$ and $y_1,\ldots,y_m$, referring to these quantities as expectation value and deviation of some preselected probability distribution function. Also one may include additional random bias if desired. Using such initial assumptions it's easy to compare numerically above listed methods. As expected from the Gauss-Markov theorem, numerical analysis gives both lowest values of the $\chi^2(f,g)$ function and the correct minimal variances $D(f)$, $D(g)$ for the last $\chi^2$ method. At same time minimum value $\chi^2(a_{+},a_{-})$ for $\chi^2$-$a$ procedure equals to the minimum $\chi^2(f,g)$ value, although presence of the random data in the mean values $m(a_{-},a_{+})$ increases the errors $D[f(a)]$, $D[g(a)]$. Both orthogonal method and optimized orthogonal method have the same larger $\chi^2$ values for different optimal sets of $a$ and $b$ and for nonoptimized $a_{\pm}$ parameters, though they give different $D(f)$, $D(g)$ values.  These estimations include bias to be calculated analytically, contrary to the linear $\chi^2$-$f,g$ procedure without it. This causes additional peculiarity -- one can get in this case either larger $D(f)$ and smaller $D(g)$ or opposite, compared to the $\chi^2$ methods. Detailed analysis can be carried out analytically using standard statistical methods after weight function definition. It should be noted that the orthogonal weight function method gives expectation values $f(a_{\pm})$ and $g(a_{\pm})$ (defined by (\ref{owf})) equal to optimized expectation values $f(a_{\pm {\rm opt}})$ and $g(a_{\pm {\rm opt}})$ (\ref{opt}), although they differs from mean $\chi^2$ $f$ and $g$ values (\ref{chif}). The orthogonal weight functions approaches with formulas (\ref{owf}) and (\ref{opt}) can be used as a approximate estimations. The last mentioned least squares procedure gives reasonable unbiased estimation, and can be used for Born cross section analysis. It should be noted that discussed methods can be easily generalized for more common case and for various specific purposes.

%
%

\end{document}